\documentclass[structabstract]{aa}
\usepackage{float}
\usepackage{lineno}
\usepackage{natbib}
\bibpunct{(}{)}{;}{a}{}{,}
\usepackage{subfigure}
\usepackage{graphicx}
\usepackage{epsfig}
\usepackage{amsmath}
\usepackage[varg]{txfonts}
\usepackage{lscape}
\usepackage{longtable}
\usepackage{txfonts}
\usepackage{color}

\newcommand{\multilineL}[1]{\begin{tabular}[b]{@{}l@{}}#1\end{tabular}}

\makeatletter

\newcommand{\Rmnum}[1]{\expandafter\@slowromancap\romannumeral #1@}
\makeatother
\begin{document}
   \title{Nonlinear force-free field extrapolation in spherical geometry: improved boundary data
   treatment applied to a SOLIS/VSM vector magnetogram }
   \author{Tilaye Tadesse\inst{1,2}, T. Wiegelmann\inst{1}, B. Inhester\inst{1}
          \and A. Pevtsov\inst{3}
          }
   \institute{Max-Planck-Institut f\"ur Sonnensystemforschung,
              Max-Planck-Strasse 2, 37191 Katlenburg-Lindau, Germany\\
              \email{tadesse@mps.mpg.de; wiegelmann@mps.mpg.de; inhester@mps.mpg.de}
   \and Addis Ababa University, College of Education, Department of Physics Education, Po.Box 1176, Addis Ababa, Ethiopia\\
   \email{tilaye.tadesse@gmail.com}
   \and National Solar Observatory, Sunspot, NM 88349, U.S.A.\\
      \email{apevtsov@nso.edu}
      }
  \date{Received 19 July 2010 / Accepted 23 November 2010}
  \abstract
     {Understanding the 3D structure of coronal magnetic field is important to understand: the onset of flares and coronal mass
     ejections, the stability of active region, and to monitor the magnetic helicity \& free magnetic energy and other
     phenomena in the solar atmosphere. Routine measurements of the solar magnetic field are mainly carried out in the photosphere.
Therefore, one has to infer the field strength in the higher layers of the solar atmosphere from the measured photospheric
field based on the assumption that the corona is force-free. Meanwhile, those measured data are inconsistent with the above
force-free assumption. Therefore, one has to apply some transformations to these data before nonlinear force-free extrapolation
codes can be applied.}
   {Extrapolation codes in cartesian geometry for modelling the magnetic field in the corona do not take the curvature of
   the Sun's surface into account and can only be applied to relatively small areas, e.g., a single active region.
     Here we apply a method for nonlinear force-free coronal magnetic field modelling and preprocessing of photospheric
     vector magnetograms in spherical geometry using the optimization procedure.
  }
   {We solve the nonlinear force-free field equations by minimizing a functional in spherical coordinates over
   a restricted area of the Sun. We extend the functional by an additional term, which allows to incorporate measurement error
   and treat regions with lacking observational data. We use vector magnetograph data from the Synoptic Optical Long-term
   Investigations of the Sun survey (SOLIS) to model the coronal magnetic field. We study two neighbouring magnetically connected
   active regions observed on May 15 2009. }
   { For vector magnetograms with variable measurement precision and randomly scattered data gaps (e.g., SOLIS/VSM)
   the new code yields field models which satisfy the solenoidal and force-free condition significantly better as it allows
   deviations between the extrapolated boundary field and observed boundary data within measurement errors. Data gaps are
   assigned to an infinite error. We extend this new scheme to spherical geometry and apply it for the first time to
   real data. }
   {}
   \keywords{Magnetic fields  -- Sun: corona -- Sun: photosphere -- methods: numerical
               }
\titlerunning{Nonlinear force-free field extrapolation of SOLIS/VSM vector magnetogram in spherical geometry}
\authorrunning {T. Tadesse \textit{et al}.}
   \maketitle
\section{Introduction}
 Observations have shown that physical conditions in the solar atmosphere are strongly controlled by
 solar magnetic field. The magnetic field also provides the link between different manifestations of
 solar activity like, for instance, sunspots, filaments, flares, or coronal mass ejections. Therefore,
 the information about the 3D structure of magnetic field vector throughout the solar atmosphere is
 crucially important. Routine measurements of the solar vector magnetic field are mainly carried out
 in the photosphere. Therefore, one has use numerical modelling to infer the field strength in the higher
 layers of the solar atmosphere from the measured photospheric field based on the assumption that the corona
 is force-free. Due to the low value of the plasma $\beta$ (the ratio of gas pressure to magnetic pressure)
 \citep{Gary}, the solar corona is magnetically dominated. To describe the equilibrium structure of the
 coronal magnetic field when non-magnetic forces are negligible, the force-free assumption is then appropriate:
\begin{equation}
   (\nabla \times\vec{B})\times\vec{B}=0 \label{a1}
\end{equation}
\begin{equation}
    \nabla \cdot\vec{B}=0 \label{a2}
 \end{equation}
\begin{equation}
    \vec{B}=\vec{H}_{obs} \quad \text{on photosphere} \label{a3}
 \end{equation}
 where $\vec{B}$ is the magnetic field and $\vec{H}_{obs}$ is 2D observed surface magnetic field on photosphere.
 Extrapolation methods have been developed for different types of force-free fields: potential field extrapolation
\citep{Schmidt,Semel67}, linear force-free field extrapolation \citep{Chiu,Seehafer,Seehafer82,Semel88,clegg00}, and
nonlinear force-free field extrapolation
\citep{Sakurai81,wu90,cuperman91,demoulin92,mikic94,Roumeliotis,Amari97,Amari99,yan00,valori05,Wheatland04,Wiegelmann04,Amari,Inhester06}.
Among these, the nonlinear force-free field has the most realistic description of the coronal magnetic field.
For a more complete review of existing methods for computing nonlinear force-free coronal magnetic fields, we refer
to the review papers by \citet{Amari97}, \citet{Schrijver06}, \cite{Metcalf}, and \citet{Wiegelmann08}.

 The magnetic field is not force-free in the photosphere, but becomes force-free roughly 400 km above the photosphere
 \citep{Metcalf:1995}. Nonlinear force-free extrapolation codes can be applied only to low plasma-$\beta$ regions, where
 the force-free assumption is justified. The preprocessing scheme as used until now modifies observed photospheric vector
 magnetograms with the aim of approximating the magnetic field vector at the bottom of the force-free
 domain \citep{Wiegelmann06sak,Fuhrmann,tilaye09}. The resulting boundary values are expected to be more
suitable for an extrapolation into a force-free field than the original values. Preprocessing is important
for those NLFF-codes which uses the magnetic field vector on the boundary directly. Consistent computations for the
Grad-Rubin method, which uses $B_{n}$ and $J_{n}$ (or $\alpha$) as boundary condition have been carried
out by \citet{Wheatland:2009}.

  In this paper, we use a larger computational domain which accommodates most of the connectivity within the coronal region.
  We also take uncertainties of measurements in vector magnetograms into accounts as suggested in \citet{DeRosa}.
  We implement a preprocessing procedure of \citet{tilaye09} to SOLIS data in spherical geometry by considering
  the curvature of the Sun's surface for the large field of view containing two active regions. We use a spherical
  version of the optimization procedure that has been implemented in cartesian geometry in \citet{Wiegelmann10}
  for synthetic boundary data.
\section{Method}
\subsection{The SOLIS/VSM instrument}

In this study, we use vector magnetogram observations from the Vector Spectromagnetograph
\citep[VSM; see][]{Jones02}, which is part of the Synoptic Optical Long-term Investigations
of the Sun (SOLIS) synoptic facility \citep[SOLIS; see][]{Keller03}. VSM/SOLIS currently operates at the
Kitt Peak National Observatory, Arizona, and it has provided magnetic field observations of the
Sun almost continuously since August 2003.

VSM is a full disk Stokes Polarimeter. As part of daily synoptic observations, it takes four different
observations in three spectral lines: Stokes $I$(intensity), $V$ (circular polarization, $Q$, and $U$
(linear polarization) in photospheric spectral lines Fe {\Rmnum{1}} 630.15 nm and Fe {\Rmnum{1}} 630.25 nm ,
Stokes $I$ and $V$ in Fe {\Rmnum{1}} 630.15 nm and Fe {\Rmnum{1}} 630.25 nm, similar observations in
chromospheric spectral line Ca {\Rmnum{2}} 854.2 nm, and Stokes $I$ in the He {\Rmnum{1}} 1083.0 nm line
and the near-by Si {\Rmnum{1}} spectral line. Observations of $I$, $Q$, $U$, and $V$ are used to construct
a full disk vector magnetograms, while $I-V$ observations are employed to create separate full disk
longitudinal magnetograms in the photosphere and the chromosphere.

In this study, we use a vector magnetogram observed on 15 May 2009. The data were taken with 1.125 arcsec
pixel size and 2.71 pm spectral sampling. ( In December 2009, SOLIS/VSM has upgraded its cameras from
Rockwell ( 90 Hz, 18 micron pixels ) to Sarnoff ( 300 Hz, 16 micron pixels ). This camera upgrade has
resulted in improved spatial and spectral sampling ). The noise level for line-of-sight component is about
1 Gauss. However, noise due to atmospheric seeing may be much larger, and the final measurement error
depends on the measured flux, its spatial distribution as well as the seeing conditions. A rough estimate
suggests a noise level of a few tens of Gauss for areas with a strong horizontal gradient of magnetic field
and about 1 arcsec atmospheric seeing.
\begin{figure*}
   \centering
 \includegraphics[bb=50 45 483 337,clip,width=1.0\textwidth]{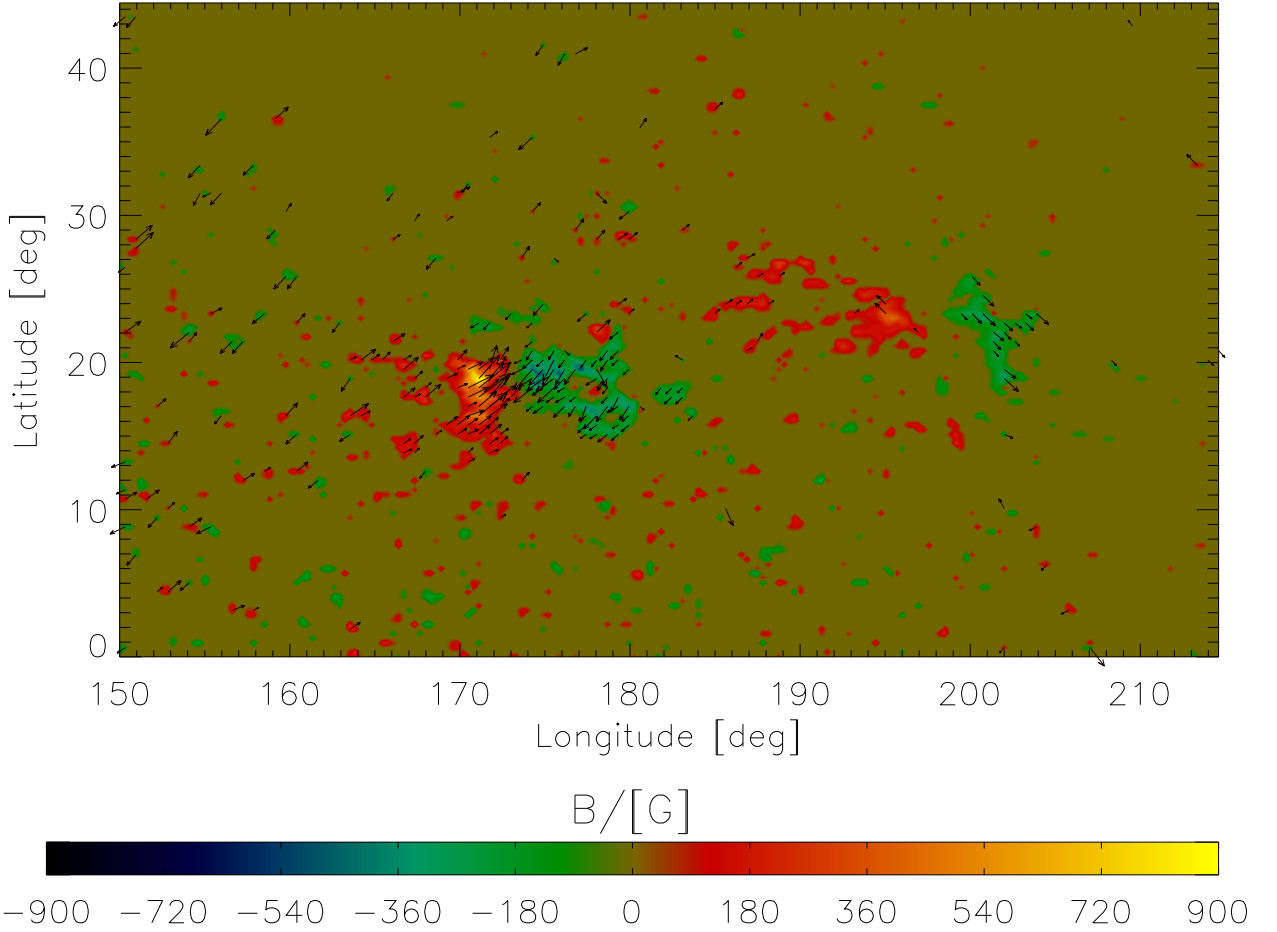}
\caption{Surface contour plot of radial magnetic field vector and vector field plot of transverse field with black arrows.}
\label{figaa}
   \end{figure*}

To create a single magnetogram, solar disk image is scanned from terrestrial South to North; it takes
about 20 minutes to complete one vector magnetogram. After the scan is done, the data are sent to an
automatic data reduction pipeline that includes dark and flat field correction. Once the spectra are
properly calibrated, full disk vector (magnetic field strength, inclination, and azimuth) magnetograms
are created using two different approaches. Quick-look (QL) vector magnetograms are created based on
algorithm by \citet{Auer:1977}. The algorithm uses the Milne-Eddington model of solar atmosphere,
which assumes that magnetic field is uniform (no gradients) through the layer of spectral line formation \citep{Unno:1956}.
It also assumes symmetric line profiles, disregards magneto-optical effects (e.g., Faraday rotation), and
does not separate contribution of magnetic and non-magnetic components in spectral line profile (i.e.,
magnetic filling factor is set to unity). A complete inversion of  spectral data is done later using
technique developed by \citet{Skumanich:1987}. This latter inversion (called ME magnetogram) also
employs Milne-Eddington model of atmosphere, but it solves for magneto-optical effects and determines
magnetic filling factor (fractional contribution of magnetic and non-magnetic components to each pixel).
The ME inversion is only performed for pixels with spectral line profiles above the noise level. For pixels
below the polarimetric noise threshold, magnetic field parameters are set to zero.

From the measurements the azimuths of transverse magnetic field can only be determined with 180-degree
ambiguity. This ambiguity is resolved using the Non-Potential Field Calculation \citep[NPFC; see][]{Georgoulis05}.
The NPFC method was selected on the basis of comparative investigation of several methods for 180-degree
ambiguity resolution \citep{Metcalf:2006}. Both QL and ME magnetograms can be used for potential and/or
force-free field extrapolation. However, in strong fields inside sunspots, the QL field strengths may
exhibit erroneous decrease inside sunspot umbra due to, so called magnetic saturation. For this study we choose
to use fully inverted ME magnetograms. Fig.~\ref{figaa} shows a map of the radial component of the field
as a contour plot with the transverse magnetic field depicted as black arrows. For this particular
dataset, about $80\%$ of the data pixels are undetermined and as a result the ratio of data gaps to total number
of pixels is large.
\subsection{Preprocessing of SOLIS data}
The preprocessing scheme of \citet{tilaye09} involves minimizing a two-dimensional functional of quadratic
form in spherical geometry as following:
 \begin{displaymath} \vec{H}=\emph{argmin}(L_{p})
\end{displaymath}
\begin{equation}
L_{p}=\mu_{1}L_{1}+\mu_{2}L_{2}+\mu_{3}L_{3}+\mu_{4}L_{4}\label{a4}
\end{equation}
where $\vec{H}$ is preprocessed surface magnetic field from the input observed field $\vec{H}_{obs}$. Each of the
constraints $L_{n}$ is weighted by a yet undetermined factor $\mu_{n}$. The first term $(n=1)$ corresponds
to the force-balance condition, the next $(n=2)$ to the torque-free condition, and the last term $(n=4)$ controls
the smoothing. The explicit form of $L_{1}$, $L_{2}$, and $L_{4}$ can be found in \citet{tilaye09}. The term
$(n=3)$ ensures that the optimized boundary condition agrees with the measured photospheric data. In the case of
SOLIS/VSM data we modified $L_{3}$ with respect to the one in \citet{tilaye09} as follows, to treat those data gaps.
\begin{equation}
 L_{3}=\sum_{p}\big(\vec{H}-\vec{H}_{obs}\big)\cdot\vec{W}(\theta,\phi)\cdot\big(\vec{H}-\vec{H}_{obs}\big)
,\label{a7}
\end{equation}
In this integral, $\vec{W}(\theta,\phi)=diag\big(w_{\text{radial}},w_{\text{trans}},w_{\text{trans}}\big)$
is a diagonal matrix which gives different weights to the different observed surface field components
depending on their relative measurement accuracy. A careful choice of the preprocessing parameters $\mu_{n}$
ensures that the preprocessed magnetic field $\vec{H}$ does not deviate from the original observed field
$\vec{H}_{obs}$ by more than the measurement errors. As the result of parameter study in this work, we found
$\mu_{1}=\mu_{2}=1.0$,  $\mu_{3}=0.03$ and $\mu_{4}=0.45$ as optimal value.
\subsection{Optimization principle}
 Equations (\ref{a1}) and (\ref{a2}) can be solved with the help of an optimization principle, as proposed
by \citet{Wheatland00} and generalized by \citet{Wiegelmann04} for cartesian geometry. The method minimizes
a joint measure of the normalized Lorentz forces and the divergence of the field throughout the volume of
interest, $V$. Throughout this minimization, the photospheric boundary of the model field $\vec{B}$ is exactly
matched to the observed $\vec{H}_{obs}$ and possibly preprocessed magnetogram values $\vec{H}$. Here, we use the optimization
approach for functional $(L_\mathrm{\omega})$ in spherical geometry \citep{Wiegelmann07,tilaye09} along with
the new method which instead of an exact match enforces a minimal deviations between the photospheric boundary
of the model field $\vec{B}$ and the magnetogram field $\vec{H}_{obs}$ by adding an appropriate surface integral term
$L_{photo}$ \citep{Wiegelmann10}.

\begin{displaymath} \vec{B}=\emph{argmin}(L_{\omega})
\end{displaymath}
\begin{equation}L_{\omega}=L_{f}+L_{d}+\nu L_{photo} \label{a9}
\end{equation}
\begin{displaymath} L_{f}=\int_{V}\omega_{f}(r,\theta,\phi)B^{-2}\big|(\nabla\times {\vec{B}})\times
{\vec{B}}\big|^2  r^2\sin\theta dr d\theta d\phi
\end{displaymath}
\begin{displaymath}L_{d}=\int_{V}\omega_{d}(r,\theta,\phi)\big|\nabla\cdot {\vec{B}}\big|^2
  r^2\sin\theta dr d\theta d\phi
\end{displaymath}
\begin{displaymath}L_{photo}=\int_{S}\big(\vec{B}-\vec{H}_{obs}\big)\cdot\vec{W}(\theta,\phi)\cdot\big(
\vec{B}-\vec{H}_{obs}\big) r^{2}\sin\theta d\theta d\phi
\end{displaymath}

  where $L_{f}$ and $L_{d}$ measure how well the force-free Eqs.~(\ref{a1}) and divergence-free (\ref{a2}) conditions
are fulfilled, respectively. $\omega_{f}(r,\theta,\phi)$ and $\omega_{d}(r,\theta,\phi)$ are weighting
functions. The third integral, $L_{photo}$, is surface integral over the photosphere which allows us to relax
the field on the photosphere towards force-free solution without to much deviation from the original surface field
data. $\vec{W}(\theta,\phi)$ is the diagonal matrix in Eq.~(\ref{a7}).

Numerical tests of the effect of the new term $L_{photo}$ were performed by \citet{Wiegelmann10} in cartesian geometry for
synthetic magnetic field vector generated from Low \& Lou model \citep{Low90}. They showed that this new method to
incorporate the observed boundary field allows to cope with data gaps as they are present in SOLIS and other vector
magnetogram data. Within this work, we use a spherical geometry for the full disk data from SOLIS. We use a spherical
grid $r$, $\theta$, $\phi$ with $n_{r}$, $n_{\theta}$, $n_{\phi}$ grid points in the direction of radius, latitude,
and longitude, respectively. The method works as follows:
\begin{itemize}
      \item We compute an initial source surface potential field in the computational domain from
$H_{robs}$, the normal component of the surface field at the photosphere at $r = 1R_\mathrm{\sun}$.
The computation is performed by assuming that a currentless ($\textbf{\textit{J}}=0$ or
$\nabla\times\textbf{\textit{B}}=0$) approximation holds between the photosphere and some spherical surface $S_{s}$
(source surface where the magnetic field vector is assumed radial). We computed the solution of this boundary-value
problem in a standard form of harmonic expansion in terms of eigen-solutions of the Laplace equation written in a spherical
coordinate system,  $(r,\theta,\phi)$.
      \item We minimize $L_{\omega}$(Eqs. \ref{a9}) iteratively without constraining $\vec{H}_{obs}$ at the photosphere
      boundary as in previous version of Wheatland algorithm \citep{Wheatland00}. The model magnetic field
      $\vec{B}$ at the surface is gradually driven towards the observations while the field in the volume $V$ relaxes
      to force-free. If the observed field is inconsistent, the difference $\vec{B}-\vec{H}_{obs}$ or $\vec{B}-\vec{H}$
      (for preprocessed data) remains finite depending in the control parameter $\nu$. At data gaps in $\vec{H}_{obs}$,
      we set $w_{radial}=0$ and  $w_{trans}=0$ and respective field value is automatically ignored.
   \item The state $L_{\omega}=0$ corresponds to a perfect force-free and divergence-free state and exact agreement of
   the boundary values $\vec{B}$ with observations $\vec{H}_{obs}$ in regions where $w_{radial}$ and $w_{trans}$ are
   greater than zero. For inconsistent boundary data the force-free and solenoidal conditions can
  still be fulfilled, but the surface term $L_{photo}$ will remain finite. This results in some deviation of the
  bottom boundary data from the observations, especially in regions where $w_{radial}$ and $w_{trans}$ are
  small. The parameter $\nu$ is tuned so that these deviations do not exceed the local estimated measurement error.
   \item The iteration stops when $L_{\omega}$ becomes stationary as $\Delta L_{\omega}/L_{\omega}<10^{-4}$.
   \end{itemize}
\section{Results}
 We use the vector magnetograph data from the Synoptic Optical Long-term Investigations of the Sun survey
 (SOLIS) to model the coronal magnetic field. We extrapolate by means of Eq.~(\ref{a9}) both the observed field
 $\vec{H}_{obs}$ measured above two active regions observed on May 15 2009 and preprocessed surface field ($\vec{H}$,
 that obtained from $\vec{H}_{obs}$ applying our preprocessing procedure). We compute 3D magnetic field in a
 wedge-shaped computational box $V$, which includes an inner physical domain $V'$ and the buffer zone (the region
 outside the physical domain), as shown in Fig. \ref{fig2} of the bottom boundary on the
 photosphere. The wedge-shaped physical domain $V'$ has its latitudinal boundaries at
 $\theta_\mathrm{min}=3^{\degr}$ and $\theta_\mathrm{max}=42^{\degr}$ , longitudinal boundaries at
 $\phi_\mathrm{min}=153^{\degr}$ and $\phi_\mathrm{max}=212^{\degr}$, and radial boundaries at
   the photosphere ($r=1R_{\sun}$) and $r=1.75R_{\sun}$.
\begin{figure*}[htp!]
   \centering
   \mbox{
       \includegraphics[bb=50 70 870 830,clip,height=5.5cm,width=6.0cm]{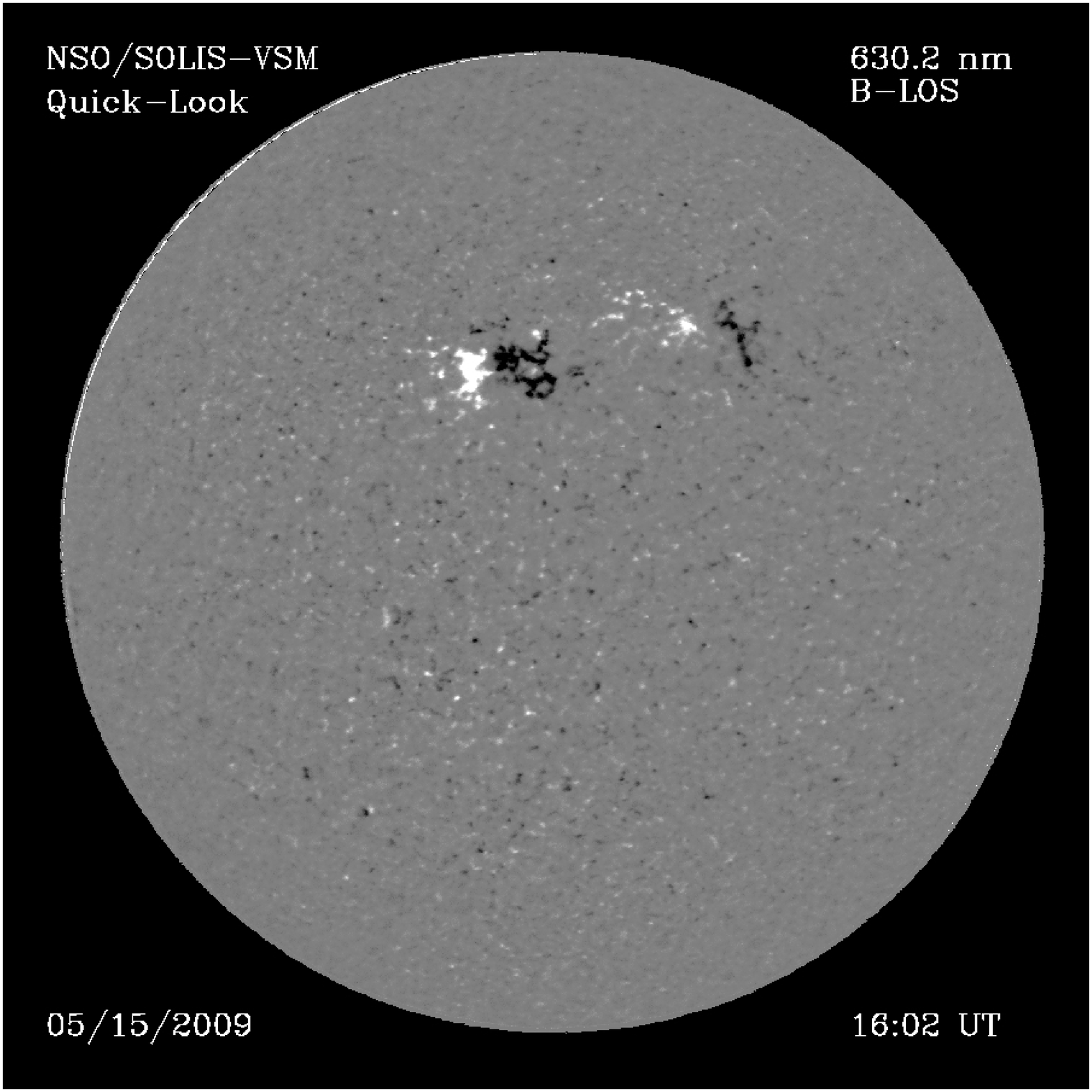}
     \includegraphics[bb=130 142 886 875,clip,height=5.5cm,width=6.0cm]{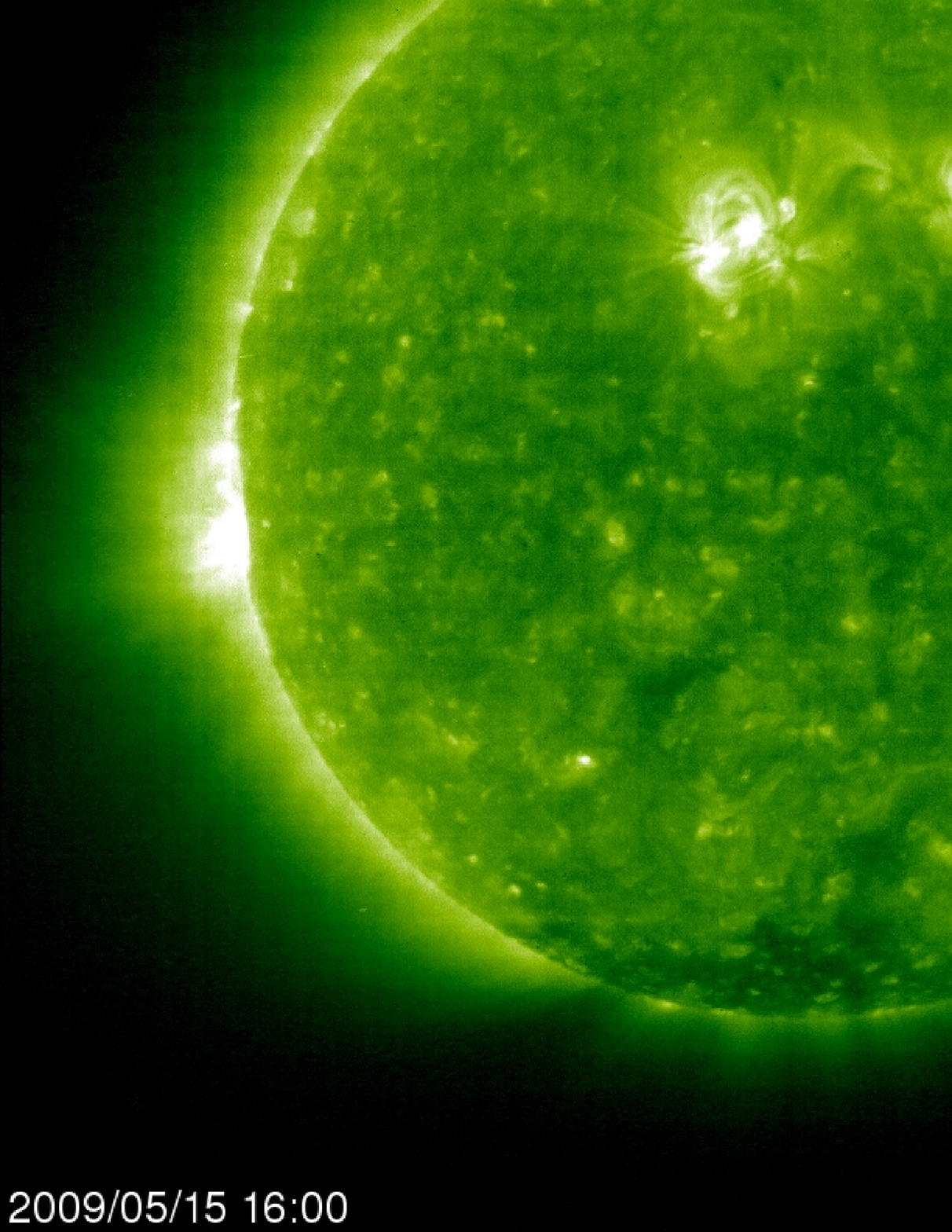}
    \includegraphics[bb=175 100 387 298,clip,height=5.5cm,width=6.0cm]{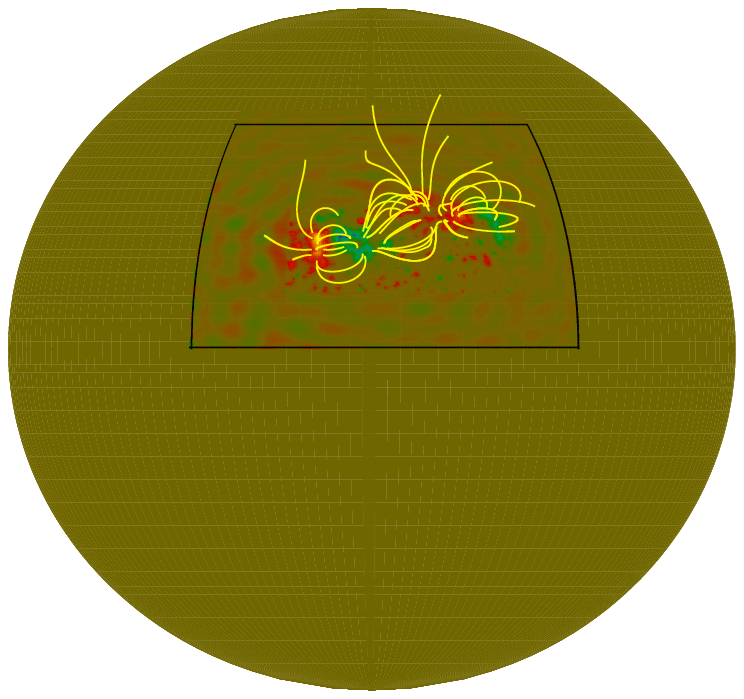}}

   \caption{ {\bf{Left:}} Full disc vector magnetogram of May 15 2009 at 16:02UT. {\bf{Middle:}} SOHO/EIT image of the
   Sun on the same day at 16:00UT. {\bf{Right:}} potential magnetic field line plot of SOLIS vector magnetogram at16:02UT, that
   has been computed from the observed radial component.}
\label{fig1}
\end{figure*}

\begin{figure*}
   \centering

   \mbox{
       \includegraphics[bb=95 105 465 354,clip,height=4.5cm,width=6.0cm]{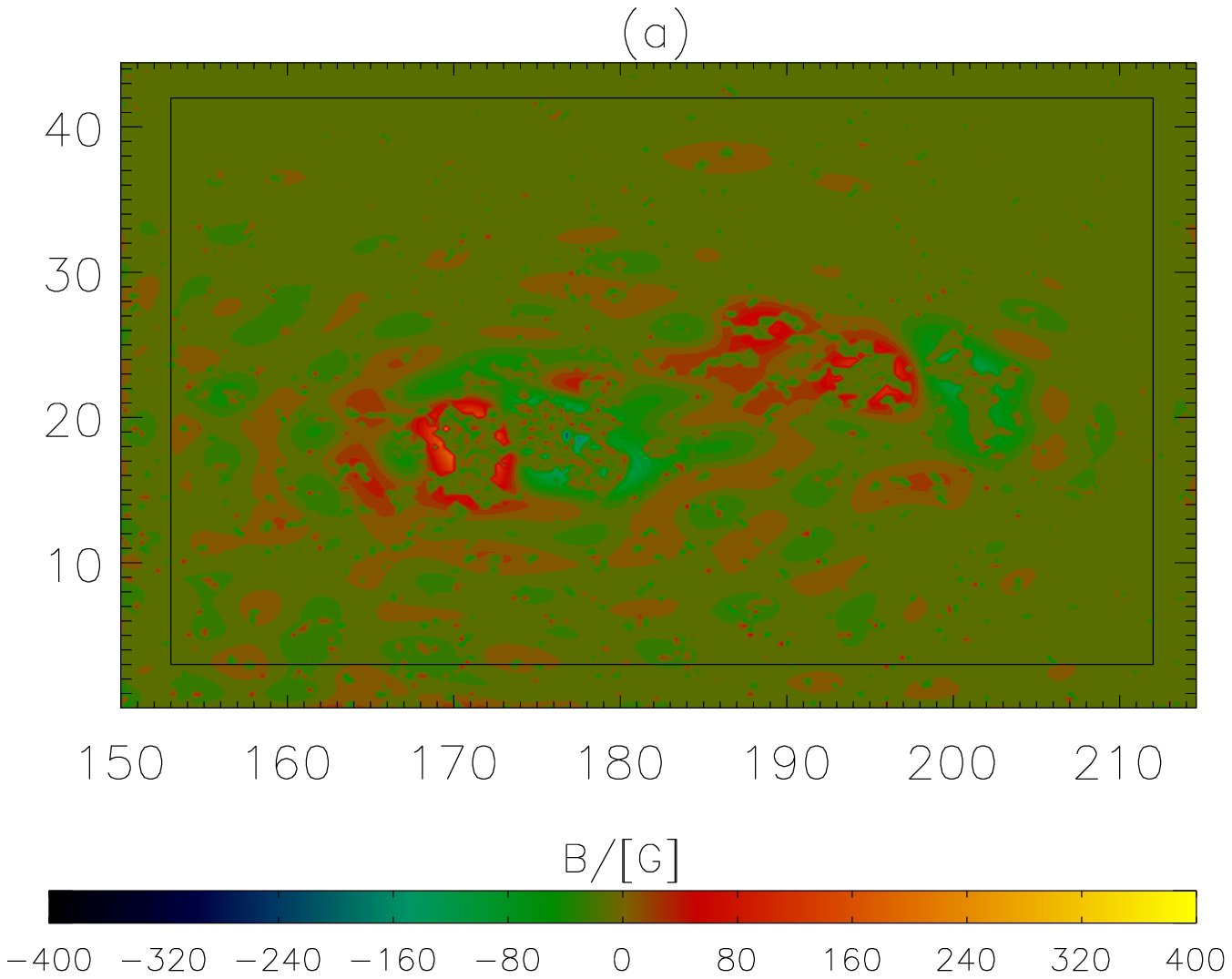}
     \includegraphics[bb=95 105 465 354,clip,height=4.5cm,width=6.0cm]{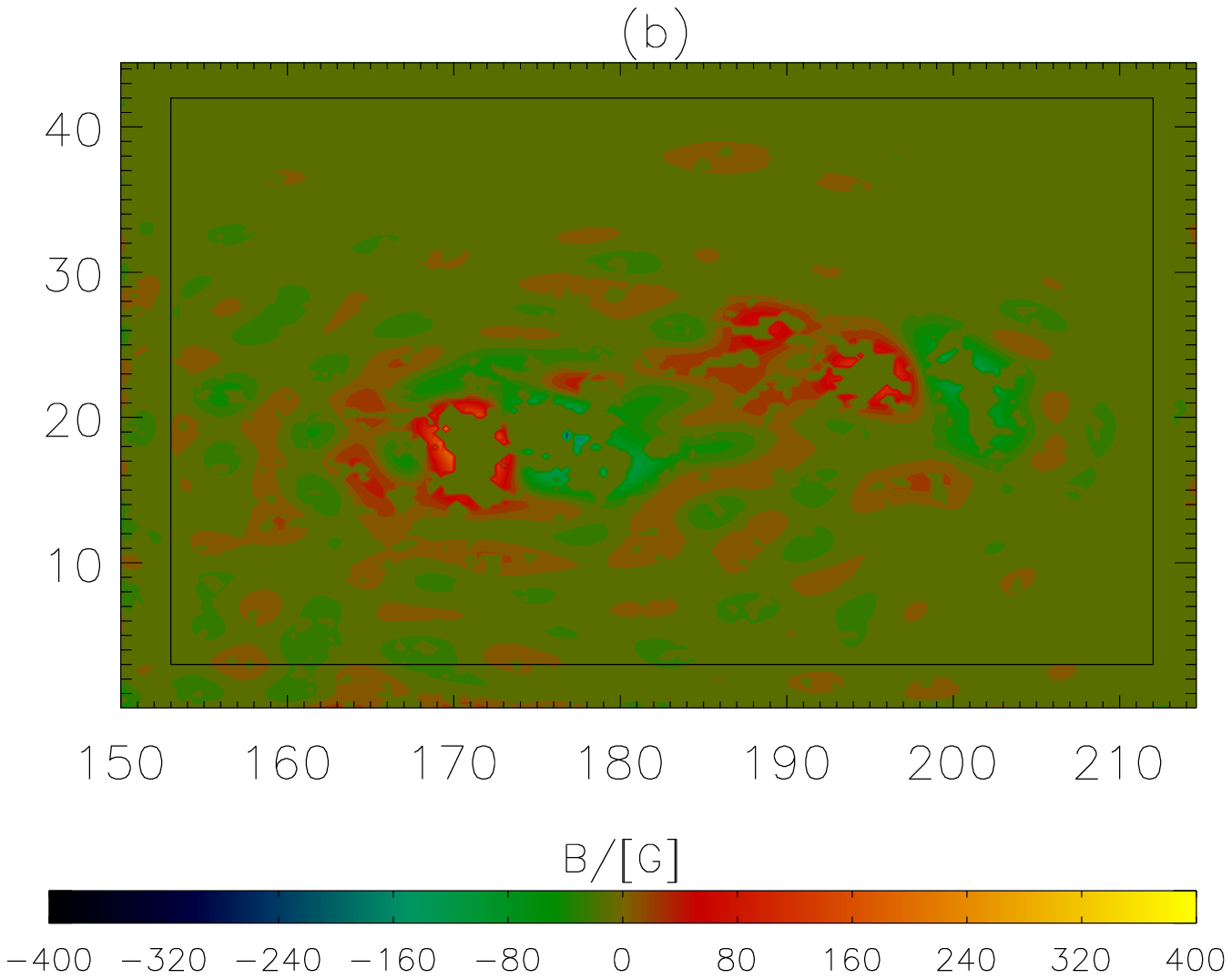}
    \includegraphics[bb=95 105 465 354,clip,height=4.5cm,width=6.0cm]{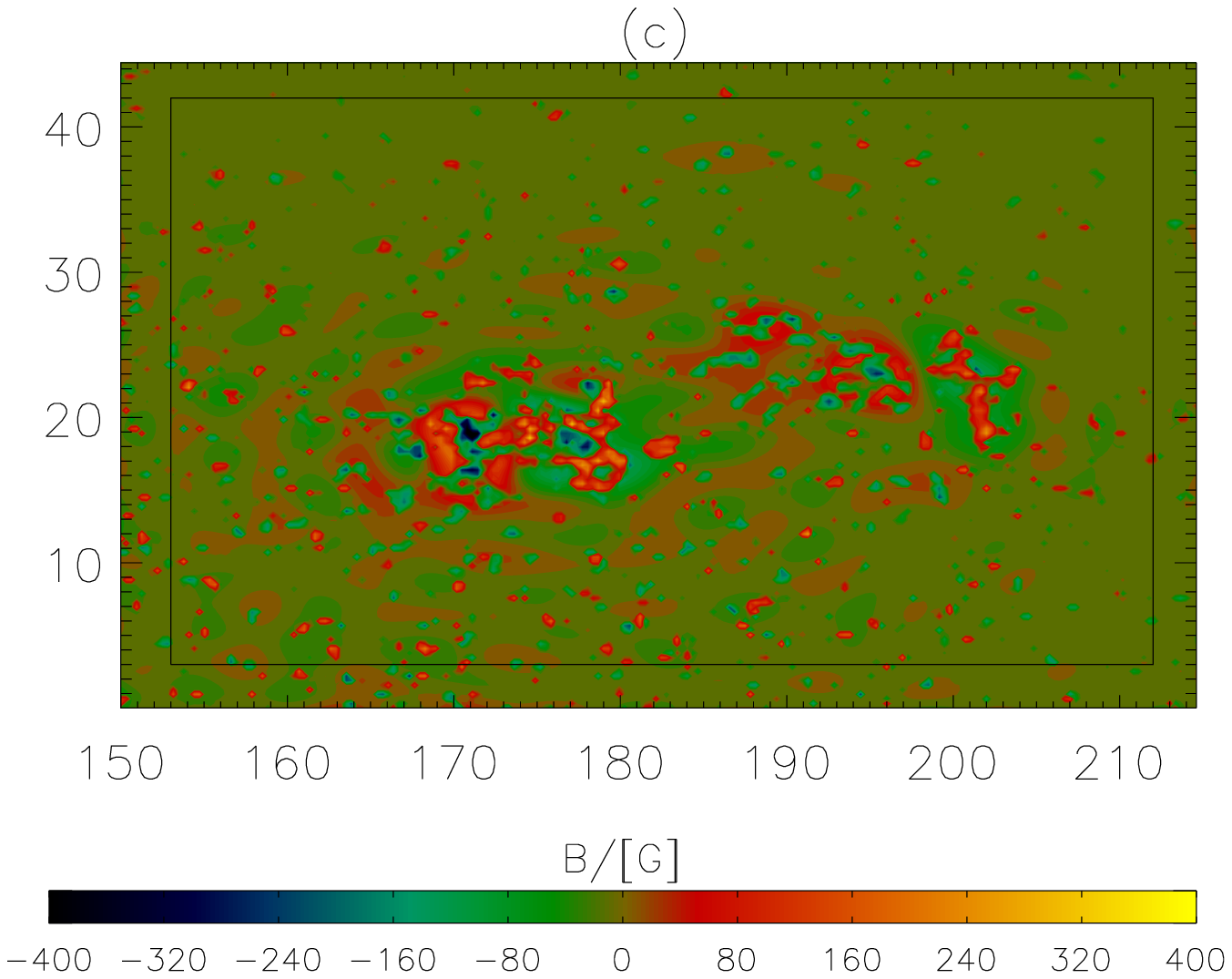}}
   \mbox{
   \includegraphics[bb=95 105 465 354,clip,height=4.5cm,width=6.0cm]{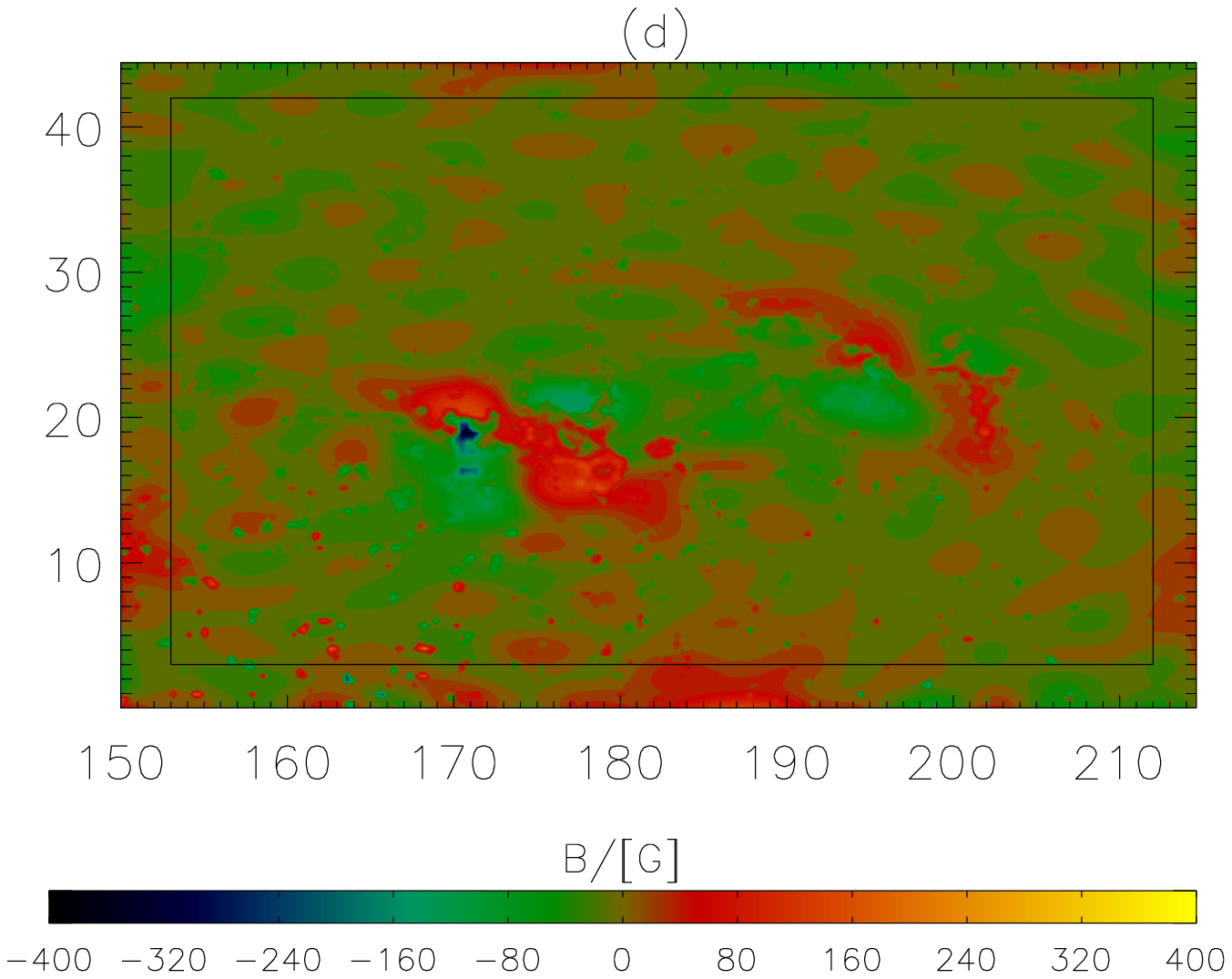}
   \includegraphics[bb=95 105 465 354,clip,height=4.5cm,width=6.0cm]{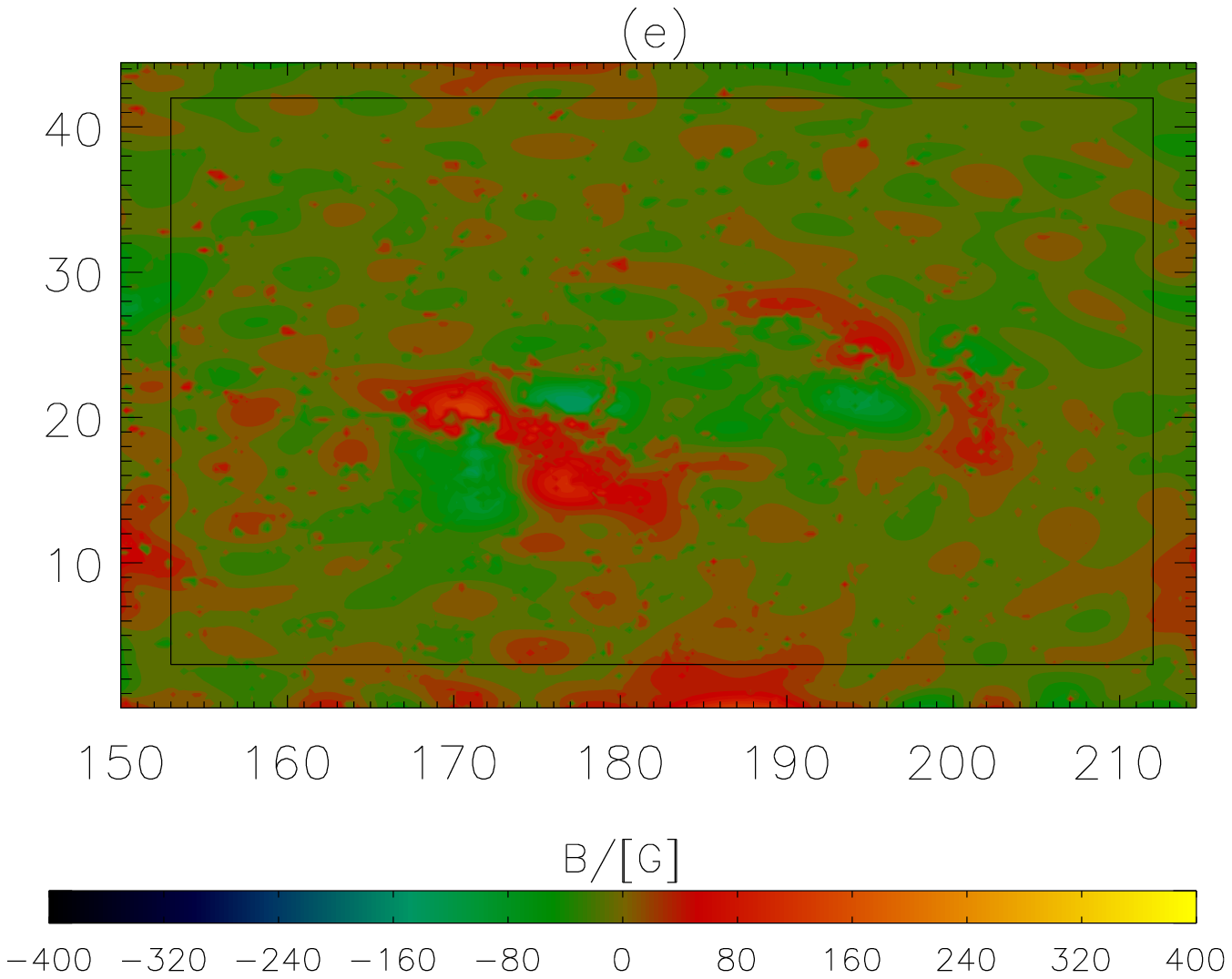}
   \includegraphics[bb=95 105 465 354,clip,height=4.5cm,width=6.0cm]{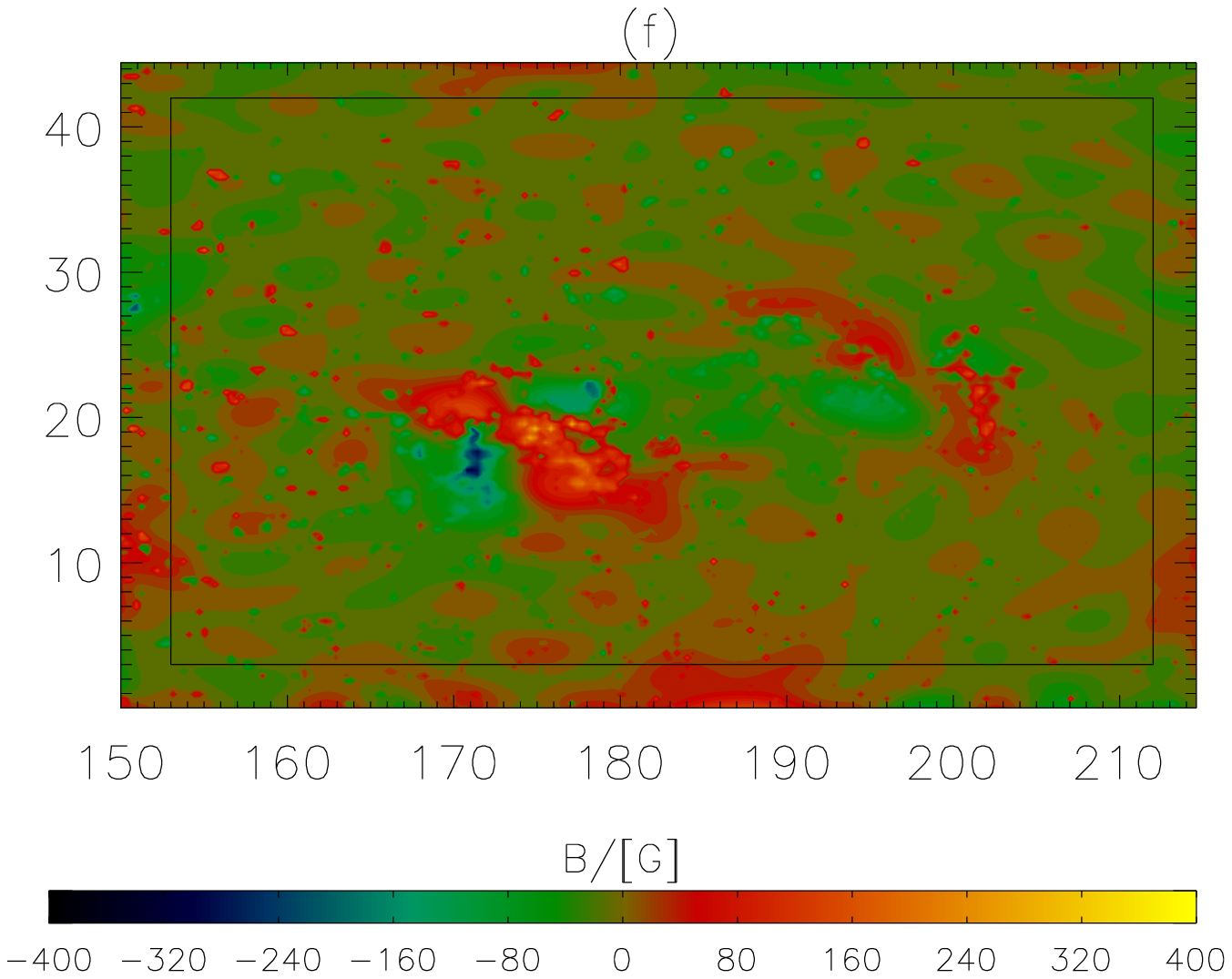}}
 \mbox{
   \includegraphics[bb=95 105 465 354,clip,height=4.5cm,width=6.0cm]{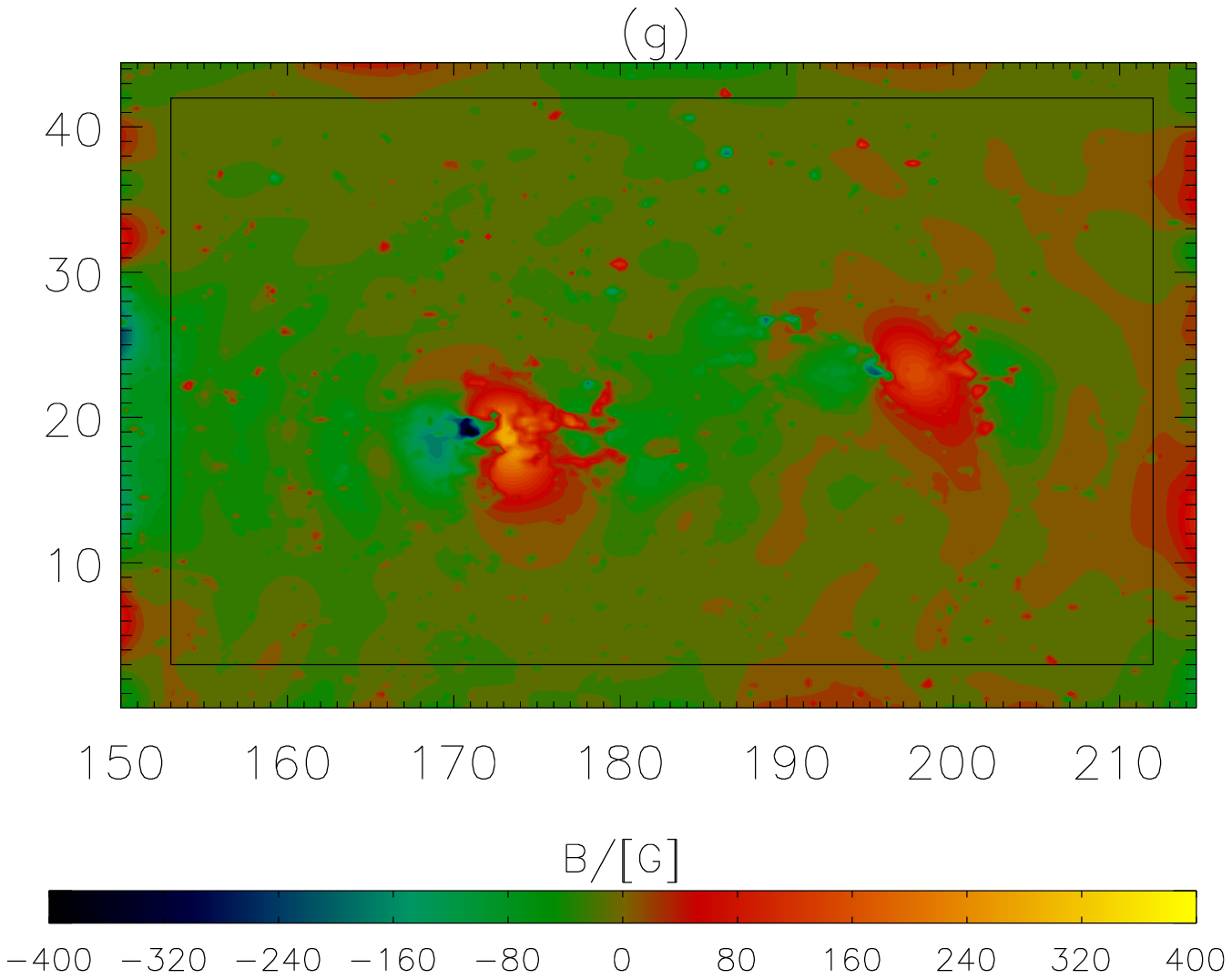}
   \includegraphics[bb=95 105 465 354,clip,height=4.5cm,width=6.0cm]{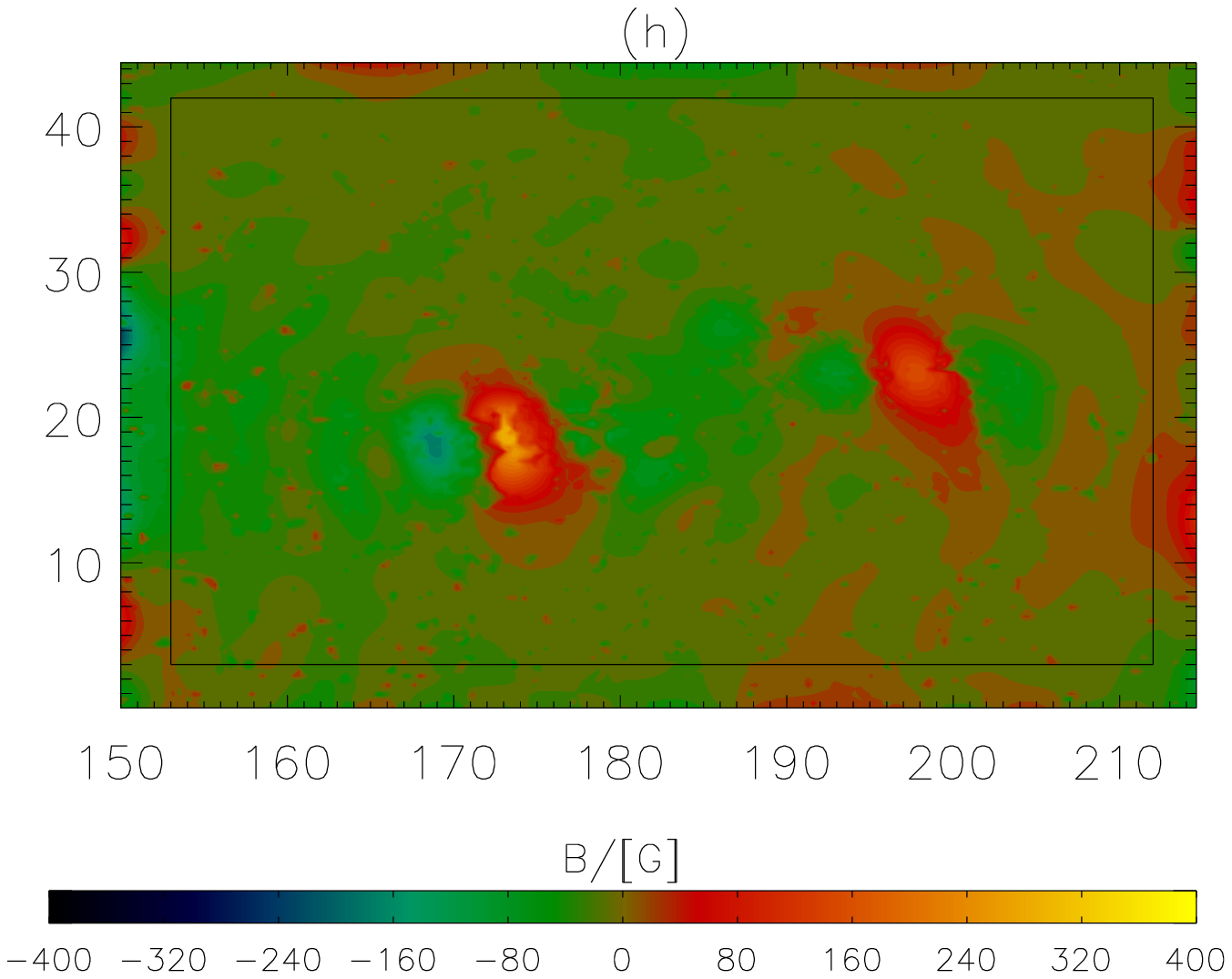}
   \includegraphics[bb=95 105 465 354,clip,height=4.5cm,width=6.0cm]{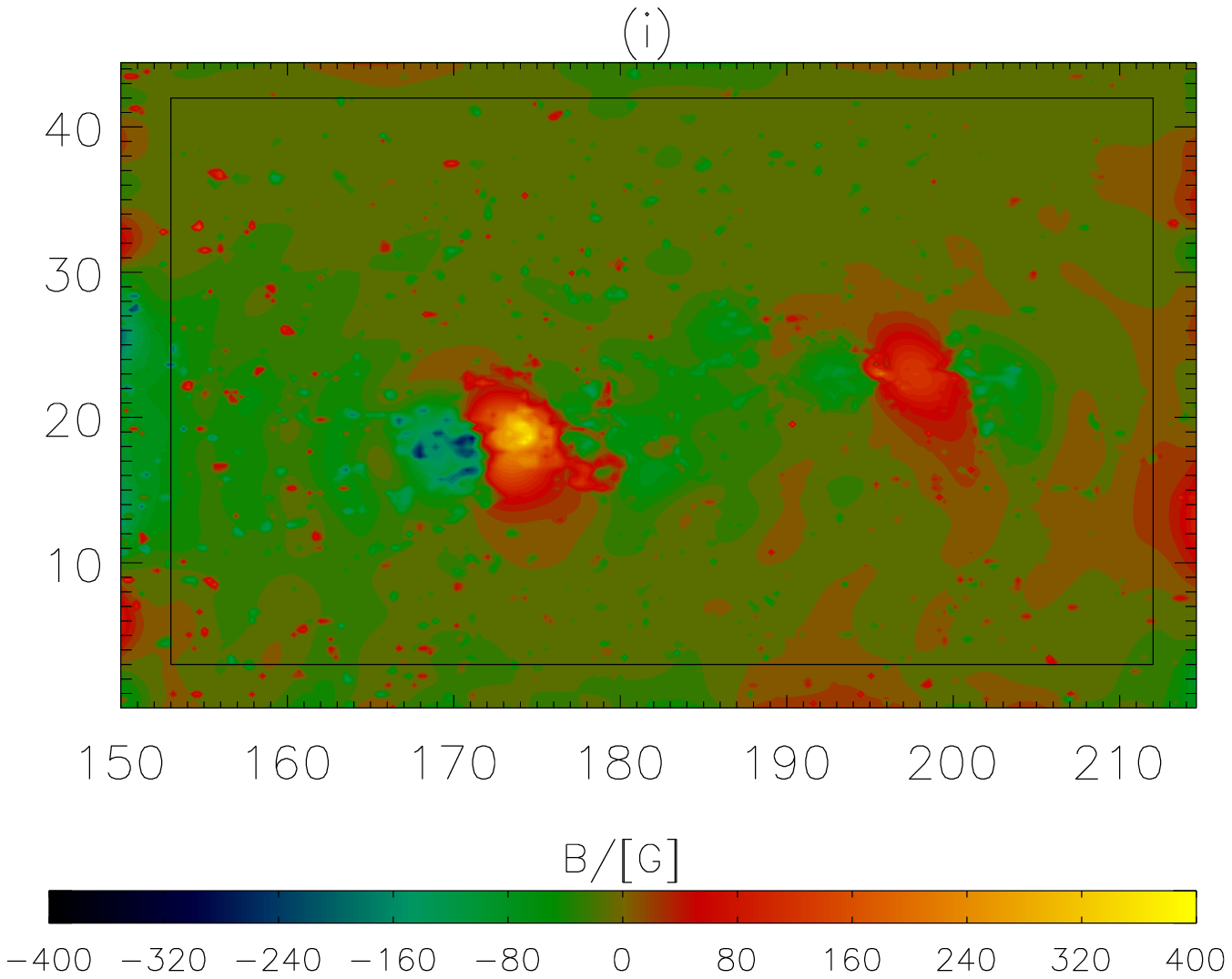}}
\begin{center}
\includegraphics[bb=82 44 478 107,clip,height=1.0cm,width=12.0cm]{15491f3c.eps}
\end{center}
   \caption{ {\bf{Top row:}} Radial surface vector field difference of a).~modelled $\vec{B}$ without preprocessing and
   $\vec{H}_{obs}$~ b).~modelled  $\vec{B}^{pre}$ and $\vec{H}_{obs}$~  c). initial potential and $\vec{H}_{obs}$.
   {\bf{Middle row:}}Latitudinal surface vector field difference of d).~modelled $\vec{B}$ without preprocessing and $\vec{H}_{obs}$~
   e).~modelled $\vec{B}^{pre}$ and $\vec{H}_{obs}$~ f).~initial potential and $\vec{H}_{obs}$.
   {\bf{Bottom row:}}Longitudinal surface vector field difference of g).~modelled $\vec{B}$ without preprocessing and $\vec{H}_{obs}$~
    h). modelled $\vec{B}^{pre}$ and $\vec{H}_{obs}$~ i).~ initial potential and $\vec{H}_{obs}$. The vertical and horizontal
    axes show latitude, $\theta$ and longitude,  $\phi$  in degree on the photosphere respectively.}
\label{fig2}
\end{figure*}
\begin{figure*}[htp!]
   \centering
  \mbox{\subfigure[Potential field]
   {\includegraphics[bb=225 189 342 270,clip,height=5.0cm,width=6.0cm]{15491f2c.eps}}
\subfigure[Field from data before preprocessing]
   {\includegraphics[bb=225 189 342 270,clip,height=5.0cm,width=6.0cm]{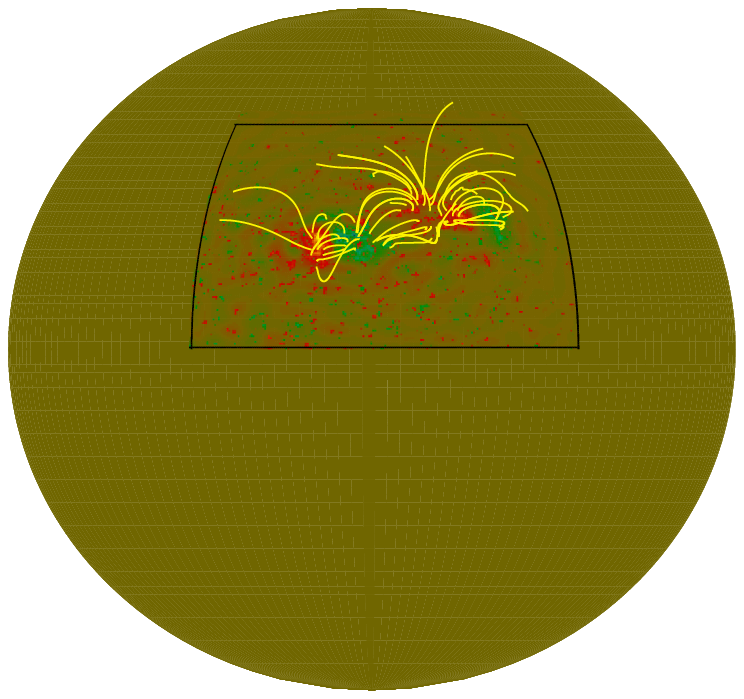}}}
\mbox{\subfigure[Field from preprocessed data]
   {\includegraphics[bb=225 189 342 270,clip,height=5.0cm,width=6.0cm]{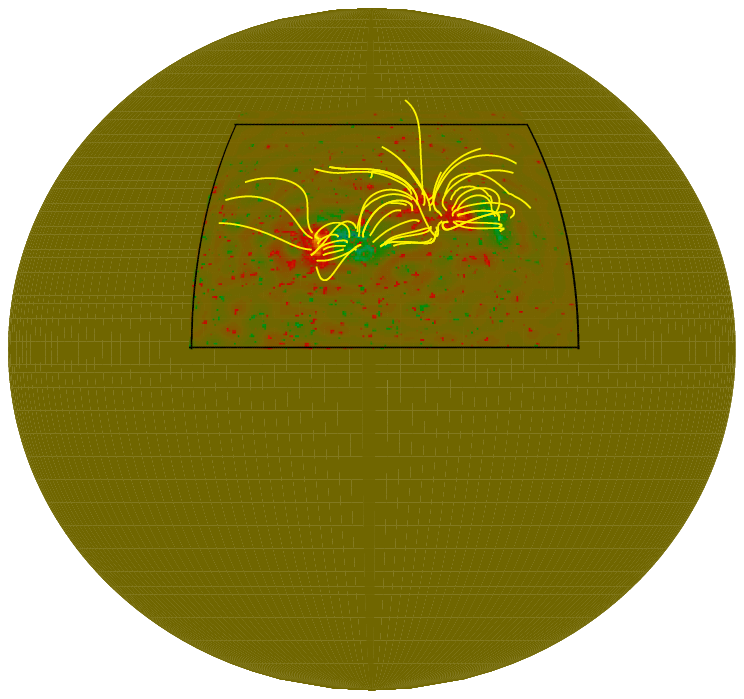}}}
     \caption{a) Some field lines for the Potential field reconstruction. b) Nonlinear force-free reconstruction
     from SOLIS data without preprocessing. c) Nonlinear force-free reconstruction from preprocessed SOLIS data. The panels
     show the same FOV as in Fig.~\ref{fig1} (right panel).}
\label{fig3}
\end{figure*}
The weighting function $\omega_{f}$ and $\omega_{d}$ in $L_{f}$ and $L_{d}$ in Eq.~(\ref{a9}) are chosen to be unity
within the inner physical domain $V'$ and decline with a cosine profile in the buffer boundary region \citep{Wiegelmann04,tilaye09}.
 They reach a zero value at the boundary of the outer volume $V$. The distance between the boundaries of $V'$ and
 $V$ is chosen to be $nd=10$ grid points wide. The framed region in Figs. \ref{fig2}.(a-i) corresponds to the
 lower boundary of the physical domain $V'$ with a resolution of $132\times 196$ pixels in the
 photosphere. The original full disc vector magnetogram has a resolution of $1788\times1788$ pixels out of
 which we extracted $142\times 206$ pixels for the lower boundary of the computational domain $V$, which
 corresponds to $550Mm\times 720Mm$ on the photosphere.

The main reason for the implementation of the new term $L_{photo}$ in Eq.~(\ref{a9}) is that we need to deal with boundary
data of different noise levels and qualities or even lack some data points completely. SOLIS/VSM provides full-disk
vector-magnetograms, but for some individual pixels the inversion from line profiles to field values may not have been
successful inverted and field data there will be missing for these pixels. Since the old code without the $L_{photo}$
term requires complete boundary information, it can not be applied to this set of SOLIS/VSM data. In our new code,
these data gaps are treated by setting $W=0$ for these pixels in Eqs.~(\ref{a9}). For those pixels, for which $\vec{H}_{obs}$
was successfully inverted, we allow deviations between the model field $\vec{B}$ and the input fields either observed
$\vec{H}_{obs}$ or preprocessed surface field $\vec{H}$ using Eqs.~(\ref{a9}) and so that the model
field can be iterated closer to a force-free solution even if the observations are inconsistent. This balance is
controlled by the Lagrangian multiplier $\nu$ as explained in \citet{Wiegelmann10}. In this work we used
$w_{radial}=100w_{trans}$ for the surface fields both from data with preprocessing and without.

Figure \ref{fig1}. shows the position of the active region on the solar disk both for SOLIS full-disk magnetogram
\footnote{http://solis.nso.edu/solis data.html}, SOHO/EIT image of the Sun observed at $195{\AA}$ on the same day at
16:00UT.\footnote{http://sohowww.nascom.nasa.gov/data/archive} As stated in section 2.3, the potential field is used as
initial condition for iterative minimization required in Eq.~\ref{a9}. The respective potential field is shown in the
rightmost panel of in Fig.~\ref{fig1}. During the iteration, the code forces the photospheric boundary of $\vec{B}$ towards
observed field values $\vec{H}_{obs}$ or $\vec{H}$ (for preprocessed data) and ignores data gaps in the magnetogram. A
deviation between surface vector field from model $\vec{B}$ and $\vec{H}_{obs}$ or $\vec{H}$ (for preprocessed data) occurs
where $\vec{H}_{obs}$ is not consistent with a force-free field. In this sense, the term $L_{photo}$ in Eq.~(\ref{a9}) acts
on $\vec{H}_{obs}$ similarly as the preprocessing, it generates a surface field $\vec{B}$ instead of $\vec{H}$ from
$\vec{H}_{obs}$ which is close to $\vec{H}_{obs}$, but consistent with a force-free field above the surface.
In Fig.~\ref{fig2} we therefore compare the option of the preprocessing and the new extrapolation code
(Eq.~\ref{a9}) on $\vec{H}_{obs}$. The figure shows the surface magnetic field differences of the preprocessed, un-preprocessed
and the potential surface fields.

In order to deternime the similarity of vector components on the bottom surface, we calculate their pixel-wise
correlations. The correlation were calculated from:
\begin{equation}
C_\mathrm{ vec}= \frac{ \sum_i \vec{v}_{i} \cdot \vec{u}_{i}}{ \Big( \sum_i |\vec{v}_{i}|^2 \sum_i
|\vec{u}_{i}|^2 \Big)^{1/2}} \label{a10}
\end{equation}
where $\vec{v}_{i}$ and $\vec{u}_{i}$ are the vectors at each grid point $i$ on the bottom surface. If the vector
fields are identical, then $C_{vec}=1$; if $\vec{v}_{i}\perp \vec{u}_{i}$ , then $C_{vec}=0$. Table \ref{table1} shows
correlations of the surface fields from $\vec{B}^{pre}-\vec{H}_{obs}$ (where $\vec{B}^{pre}$ is the model field
obtained from preprocessed surface field $\vec{H}$ using Eq.~(\ref{a9})) and $\vec{B}^{unpre}-\vec{H}_{obs}$
(where $\vec{B}^{unpre}$ is the model field obtained from observed surface field $\vec{H}_{obs}$ using Eq.~(\ref{a9})).
We have computed the vector correlations of the two surface vector fields for the three components at each grid
points to compare how well they are aligned along each directions. From those values in Table \ref{table1}
one can see that the preprocessing and extrapolation with Eq.~(\ref{a9}) act on $\vec{H}_{obs}$ in a similar way.
\begin{table}
\caption{The correlations between the components of surface fields from $(\vec{B}^{pre}-\vec{H}_{obs})$ and
$(\vec{B}^{unpre}-\vec{H}_{obs})$.}
\label{table1}
\begin{tabular}{ccc}
  \hline \hline
v & u & $C_\mathrm{vec}$\\
\hline

$(\vec{B}^{unpre}-\vec{H}_{obs})_{r}$&$ (\vec{B}^{pre}-\vec{H}_{obs})_{r}$ &$0.930$\\
 $(\vec{B}^{unpre}-\vec{H}_{obs})_{\theta}$&$ (\vec{B}^{pre}-\vec{H}_{obs})_{\theta}$ &$0.897$\\
$(\vec{B}^{unpre}-\vec{H}_{obs})_{\phi} $&$(\vec{B}^{pre}-\vec{H}_{obs})_{\phi}$ &$0.875$\\
\hline
\end{tabular}
\end{table}
In Fig.\ref{fig3}. we plot magnetic field lines for the three configurations in which the vector correlations
of potential field lines in 3D box to both the extrapolated NLFF with and without preprocessing data are $0.741$ and $0.793$,
respectively.

To understand the physics of solar flares, including the local reorganization of the magnetic field and the
acceleration of energetic particles, one has to estimate the free magnetic energy available for such
phenomena. this is the free energy which can be converted into kinetic and thermal energy. From the energy
budget and the observed magnetic activity in the active region, \citet{Regnier} and \citet{Thalmann} have investigated
the free energy above the minimum-energy state for the flare process. We estimate the free magnetic energy
from the difference of the extrapolated force-free fields and the potential field with the same normal
boundary conditions in the photosphere. We therefore estimate the upper limit to the free magnetic energy associated
with coronal currents of the form
\begin{equation}
E_\mathrm{free}=\frac{1}{8\pi}\int_{V}\Big(B_{nlff}^{2}-B_{pot}^{2}\Big)r^{2}sin\theta dr d\theta d\phi \label{ten}
\end{equation}

\begin{center}
\begin{table}
\caption{The magnetic energy associated with extrapolated NLFF field configurations with and without preprocessing.}
\label{table2}
\begin{tabular}{ccc}
 \hline \hline
Model & $E_{nlff}(10^{32}erg)$& $E_\mathrm{free}(10^{32}erg)$\\
\hline

No preprocessing&$37.456$&$4.915$\\
Preprocessed&$37.341$&$4.800$\\
\hline
\end{tabular}
\end{table}
\end{center}
where $B_{pot}$ and $B_{nlff}$ represent the potential and NLFF magnetic field,
respectively. The free energy is about $5\times10^{32}erg$. The magnetic energy associated with the
potential field configuration is found to be $32.541\times10^{32}erg$. Hence $E_{nlff}$ exceeds $E_{pot}$ by only 15$\%$.
Table \ref{table2} shows the magnetic energy associated with extrapolated NLFF field
configurations with and without preprocessing. The magnetic energy of the NLFF field
configuration obtained from the data without preprocessing is a slightly large from the
preprocessed boundary field, as the preprocessing procedure removes small scale structures.

\begin{figure}
   \centering
  \includegraphics[bb=163 240 615 470,clip,height=3.5cm,width=7.5cm]{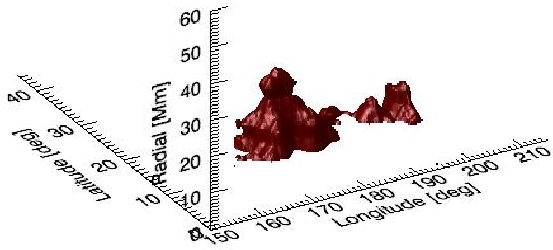}
\caption{Iso-surfaces (ISs) of the absolute current density vector $|J|=100mA\cdot m^{-2} $ computed above the active regions.
}
\label{fig4}
   \end{figure}

The electric current density calculated from Amp\`{e}re's law, $J=\triangledown\times\vec{B}/4\pi$, on
the basis of spatially sampled transverse magnetic fields varies widely over an active region. In order to
investigate how errors in the vector magnetograph measurements produce errors in the vertical electric current
densities, \citet{Liang} have numerically simulated the effects of random noise on a standard photospheric
magnetic configuration produced by electric currents satisfying the force-free field conditions. Even if the
current density can be estimated on the photosphere, it is not intuitively clear how the change in the current
density distribution affects a coronal magnetic configuration. \citet{Regnier07} studied such modifications in
terms of the geometry of field lines, the storage of magnetic energy and the amount of magnetic helicity.
Fig.\ref{fig4}. shows Iso-surface plots of current density above the volume of the active region. There are
strong current configurations above each active regions. This becomes clear if we compare the total
current in between each active region with the current from the left to the right active region. These currents
were added up from the surface normal currents emanating from those pixels which are magnetically connected inside
or across active regions respectively. The result is shown in Table~\ref{table3a}. The active regions share
a decent amount of magnetic flux compared to their internal flux from one polarity to the other. In terms
of the electric current they are much more isolated. The ratio of shared to the intrinsic magnetic flux is order
of unity, while for the electric current those ratios are much less, $1.58/49.6$ and $1.58/32.17$, respectively.
Similarly we can calculate the average value of $\alpha$ on the field lines with the respective magnetic connectivity.
The averages are shown in the second row of Table~\ref{table3a}. The two active regions are magnetically connected
but much less by electric currents.
\begin{center}
\begin{table}
\caption{The currents and average $\alpha$ calculated from those pixels which are magnetically connected. The
currents are given in Amp\`{e}re (A). }
\label{table3a}
\begin{tabular}{cccc}
 \hline \hline
 &\multilineL{Inside left\\active region} &\multilineL{Between left\\ and
 right ARs}& \multilineL{Inside right\\ active region} \\
\hline

Magnetic flux $(10^{19}Gcm^{2})$&$3.32$&$4.61$&$2.08$\\
Total current~($10^{6}$A)&$49.6$&$1.58$&$32.17$\\
Average $\alpha$~$(Mm^{-1})$&$2.49$&$0.08$&$1.62$\\
\hline
\end{tabular}
\end{table}
\end{center}
\section{Conclusion and outlook}
We have investigated the coronal magnetic field associated with the AR 11017 on 2009 May 15 and neighbouring active region
by analysing SOLIS/VSM data. We have used the optimization method for the reconstruction of nonlinear
force-free coronal magnetic fields in spherical geometry by restricting the code to limited parts of the Sun
\citep{Wiegelmann07,tilaye09}. Different from previous implementations our new code allows us to deal with lacking data and
regions with poor signal-to-noise ratio in the extrapolation in a systematic manner because it produces a field which is closer
 to a force-free and divergence-free field and tries to match the boundary only where it has been reliably
 measured \citep{Wiegelmann10}.

 For vector magnetograms with lacking data point where zero values have been replaced for
the signal below certain threshold value, the new code relaxes the boundary and allows to fulfill the solenoidal and force-free
condition significantly better as it allows deviations between the extrapolated boundary field and inconsistent observed boundary
data. With the new term $L_{photo}$ extrapolation from $\vec{H}_{obs}$ and $\vec{H}$ yield slightly the same 3D field. However,
in the latter case the iteration to minimize Eq.~(\ref{a9}) saturates in fewer iteration steps. At the same time, preprocessing
does not affect the overall configuration of magnetic field and its total energy content.

We plan to use this newly developed code for upcoming data from SDO (Solar Dynamics Observatory)/HMI ( Helioseismic and
Magnetic Imager) when full disc magnetogram data become available.
\begin{acknowledgements} SOLIS/VSM vector magnetograms are produced cooperatively by NSF/NSO and NASA/LWS. The National Solar
Observatory (NSO) is operated by the Association of Universities for Research in Astronomy,
Inc., under cooperative agreement with the National Science Foundation. Tilaye Tadesse acknowledges a fellowship of
the International Max-Planck Research School at the Max-Planck Institute for Solar System Research and the work of T.
Wiegelmann was supported by DLR-grant $50$ OC $453$  $0501$.
\end{acknowledgements}
\bibliographystyle{aa}
\bibliography{15491_bibtex}

\begin{thebibliography}{46}
\expandafter\ifx\csname natexlab\endcsname\relax\def\natexlab#1{#1}\fi

\bibitem[{{Amari} {et~al.}(1997){Amari}, {Aly}, {Luciani}, {Boulmezaoud}, \&
  {Mikic}}]{Amari97}
{Amari}, T., {Aly}, J.~J., {Luciani}, J.~F., {Boulmezaoud}, T.~Z., \& {Mikic},
  Z. 1997, Sol. Phys., 174, 129

\bibitem[{{Amari} {et~al.}(2006){Amari}, {Boulmezaoud}, \& {Aly}}]{Amari}
{Amari}, T., {Boulmezaoud}, T.~Z., \& {Aly}, J.~J. 2006, A\&A, 446, 691

\bibitem[{{Amari} {et~al.}(1999){Amari}, {Boulmezaoud}, \& {Mikic}}]{Amari99}
{Amari}, T., {Boulmezaoud}, T.~Z., \& {Mikic}, Z. 1999, A\&A, 350, 1051

\bibitem[{{Auer} {et~al.}(1977){Auer}, {House}, \& {Heasley}}]{Auer:1977}
{Auer}, L.~H., {House}, L.~L., \& {Heasley}, J.~N. 1977, Sol. Phys., 55, 47

\bibitem[{{Chiu} \& {Hilton}(1977)}]{Chiu}
{Chiu}, Y.~T. \& {Hilton}, H.~H. 1977, ApJ, 212, 873

\bibitem[{{Clegg} {et~al.}(2000){Clegg}, {Browning}, {Laurence}, {Bromage}, \&
  {Stredulinsky}}]{clegg00}
{Clegg}, J.~R., {Browning}, P.~K., {Laurence}, P., {Bromage}, B.~J.~I., \&
  {Stredulinsky}, E. 2000, \aap, 361, 743

\bibitem[{{Cuperman} {et~al.}(1991){Cuperman}, {Demoulin}, \&
  {Semel}}]{cuperman91}
{Cuperman}, S., {Demoulin}, P., \& {Semel}, M. 1991, \aap, 245, 285

\bibitem[{{Demoulin} {et~al.}(1992){Demoulin}, {Cuperman}, \&
  {Semel}}]{demoulin92}
{Demoulin}, P., {Cuperman}, S., \& {Semel}, M. 1992, \aap, 263, 351

\bibitem[{{DeRosa} {et~al.}(2009){DeRosa}, {Schrijver}, {Barnes}, {Leka},
  {Lites}, {Aschwanden}, {Amari}, {Canou}, {McTiernan}, {R{\'e}gnier},
  {Thalmann}, {Valori}, {Wheatland}, {Wiegelmann}, {Cheung}, {Conlon},
  {Fuhrmann}, {Inhester}, \& {Tadesse}}]{DeRosa}
{DeRosa}, M.~L., {Schrijver}, C.~J., {Barnes}, G., {et~al.} 2009, \apj, 696,
  1780

\bibitem[{{Fuhrmann} {et~al.}(2007){Fuhrmann}, {Seehafer}, \&
  {Valori}}]{Fuhrmann}
{Fuhrmann}, M., {Seehafer}, N., \& {Valori}, G. 2007, \aap, 476, 349

\bibitem[{{Gary}(2001)}]{Gary}
{Gary}, G.~A. 2001, Sol. Phys., 203, 71

\bibitem[{{Georgoulis}(2005)}]{Georgoulis05}
{Georgoulis}, M.~K. 2005, ApJL, 629, L69

\bibitem[{{Inhester} \& {Wiegelmann}(2006)}]{Inhester06}
{Inhester}, B. \& {Wiegelmann}, T. 2006, Sol. Phys., 235, 201

\bibitem[{{Jones} {et~al.}(2002){Jones}, {Harvey}, {Henney}, {Hill}, \&
  {Keller}}]{Jones02}
{Jones}, H.~P., {Harvey}, J.~W., {Henney}, C.~J., {Hill}, F., \& {Keller},
  U.~C. 2002, ESA SP, 505, 15

\bibitem[{{Keller} {et~al.}(2003){Keller}, {Harvey}, \& {Giampapa}}]{Keller03}
{Keller}, U.~C., {Harvey}, J.~W., \& {Giampapa}, M.~S. 2003, 4853, 194

\bibitem[{{Liang} {et~al.}(2009){Liang}, {Ma}, {Zhao}, \& {Xiang}}]{Liang}
{Liang}, H.~F., {Ma}, L., {Zhao}, H.~J., \& {Xiang}, F.~Y. 2009, New Astronomy,
  14, 294

\bibitem[{{Low} \& {Lou}(1990)}]{Low90}
{Low}, B.~C. \& {Lou}, Y.~Q. 1990, ApJ, 352, 343

\bibitem[{{Metcalf} {et~al.}(2008){Metcalf}, {Derosa}, {Schrijver}, {Barnes},
  {van Ballegooijen}, {Wiegelmann}, {Wheatland}, {Valori}, \&
  {McTtiernan}}]{Metcalf}
{Metcalf}, T.~R., {Derosa}, M.~L., {Schrijver}, C.~J., {et~al.} 2008, \solphys,
  247, 269

\bibitem[{{Metcalf} {et~al.}(1995){Metcalf}, {Jiao}, {McClymont}, {Canfield},
  \& {Uitenbroek}}]{Metcalf:1995}
{Metcalf}, T.~R., {Jiao}, L., {McClymont}, A.~N., {Canfield}, R.~C., \&
  {Uitenbroek}, H. 1995, \apj, 439, 474

\bibitem[{{Metcalf} {et~al.}(2006){Metcalf}, {Leka}, {Barnes}, {Lites},
  {Georgoulis}, {Pevtsov}, {Balasubramaniam}, {Gary}, {Jing}, {Li}, {Liu},
  {Wang}, {Abramenko}, {Yurchyshyn}, \& {Moon}}]{Metcalf:2006}
{Metcalf}, T.~R., {Leka}, K.~D., {Barnes}, G., {et~al.} 2006, Sol. Phys., 237,
  267

\bibitem[{{Mikic} \& {McClymont}(1994)}]{mikic94}
{Mikic}, Z. \& {McClymont}, A.~N. 1994, in Astronomical Society of the Pacific
  Conference Series, Vol.~68, Solar Active Region Evolution: Comparing Models
  with Observations, ed. K.~S. {Balasubramaniam} \& G.~W. {Simon}, 225--+

\bibitem[{{R{\'e}gnier} \& {Priest}(2007{\natexlab{a}})}]{Regnier}
{R{\'e}gnier}, S. \& {Priest}, E.~R. 2007{\natexlab{a}}, \apjl, 669, L53

\bibitem[{{R{\'e}gnier} \& {Priest}(2007{\natexlab{b}})}]{Regnier07}
{R{\'e}gnier}, S. \& {Priest}, E.~R. 2007{\natexlab{b}}, \aap, 468, 701

\bibitem[{{Roumeliotis}(1996)}]{Roumeliotis}
{Roumeliotis}, G. 1996, ApJ, 473, 1095

\bibitem[{{Sakurai}(1981)}]{Sakurai81}
{Sakurai}, T. 1981, Sol. Phys., 69, 343

\bibitem[{{Schmidt}(1964)}]{Schmidt}
{Schmidt}, H.~U. 1964, in The Physics of Solar Flares, 107--+

\bibitem[{{Schrijver} {et~al.}(2006){Schrijver}, {Derosa}, {Metcalf}, {Liu},
  {McTiernan}, {R{\'e}gnier}, {Valori}, {Wheatland}, \&
  {Wiegelmann}}]{Schrijver06}
{Schrijver}, C.~J., {Derosa}, M.~L., {Metcalf}, T.~R., {et~al.} 2006, Sol.
  Phys., 235, 161

\bibitem[{{Seehafer}(1978)}]{Seehafer}
{Seehafer}, N. 1978, Sol. Phys., 58, 215

\bibitem[{{Seehafer}(1982)}]{Seehafer82}
{Seehafer}, N. 1982, Sol. Phys., 81, 69

\bibitem[{{Semel}(1967)}]{Semel67}
{Semel}, M. 1967, Annales d'Astrophysique, 30, 513

\bibitem[{{Semel}(1988)}]{Semel88}
{Semel}, M. 1988, A\&A, 198, 293

\bibitem[{{Skumanich} \& {Lites}(1987)}]{Skumanich:1987}
{Skumanich}, A. \& {Lites}, B.~W. 1987, \apj, 322, 473

\bibitem[{{Tadesse} {et~al.}(2009){Tadesse}, {Wiegelmann}, \&
  {Inhester}}]{tilaye09}
{Tadesse}, T., {Wiegelmann}, T., \& {Inhester}, B. 2009, \aap, 508, 421

\bibitem[{{Thalmann} {et~al.}(2008){Thalmann}, {Wiegelmann}, \&
  {Raouafi}}]{Thalmann}
{Thalmann}, J.~K., {Wiegelmann}, T., \& {Raouafi}, N.-E. 2008, \aap, 488, L71

\bibitem[{{Unno}(1956)}]{Unno:1956}
{Unno}, W. 1956, Publ. Astron. Soc. Japan, 8, 108

\bibitem[{{Valori} {et~al.}(2005){Valori}, {Kliem}, \& {Keppens}}]{valori05}
{Valori}, G., {Kliem}, B., \& {Keppens}, R. 2005, \aap, 433, 335

\bibitem[{{Wheatland}(2004)}]{Wheatland04}
{Wheatland}, M.~S. 2004, \solphys, 222, 247

\bibitem[{{Wheatland} \& {R{\'e}gnier}(2009)}]{Wheatland:2009}
{Wheatland}, M.~S. \& {R{\'e}gnier}, S. 2009, \apjl, 700, L88

\bibitem[{{Wheatland} {et~al.}(2000){Wheatland}, {Sturrock}, \&
  {Roumeliotis}}]{Wheatland00}
{Wheatland}, M.~S., {Sturrock}, P.~A., \& {Roumeliotis}, G. 2000, ApJ, 540,
  1150

\bibitem[{{Wiegelmann}(2004)}]{Wiegelmann04}
{Wiegelmann}, T. 2004, Sol. Phys., 219, 87

\bibitem[{{Wiegelmann}(2007)}]{Wiegelmann07}
{Wiegelmann}, T. 2007, Sol. Phys., 240, 227

\bibitem[{{Wiegelmann}(2008)}]{Wiegelmann08}
{Wiegelmann}, T. 2008, Journal of Geophysical Research (Space Physics), 113, 3

\bibitem[{{Wiegelmann} \& {Inhester}(2010)}]{Wiegelmann10}
{Wiegelmann}, T. \& {Inhester}, B. 2010, A\&A, 516, A107+

\bibitem[{{Wiegelmann} {et~al.}(2006){Wiegelmann}, {Inhester}, \&
  {Sakurai}}]{Wiegelmann06sak}
{Wiegelmann}, T., {Inhester}, B., \& {Sakurai}, T. 2006, Sol. Phys., 233, 215

\bibitem[{{Wu} {et~al.}(1990){Wu}, {Sun}, {Chang}, {Hagyard}, \& {Gary}}]{wu90}
{Wu}, S.~T., {Sun}, M.~T., {Chang}, H.~M., {Hagyard}, M.~J., \& {Gary}, G.~A.
  1990, \apj, 362, 698

\bibitem[{{Yan} \& {Sakurai}(2000)}]{yan00}
{Yan}, Y. \& {Sakurai}, T. 2000, \solphys, 195, 89

\end{thebibliography}

\end{document}